\documentclass[twocolumn,showpacs,preprintnumbers,amsmath,amssymb]{revtex4}
\usepackage{amsmath}
\usepackage{graphicx}
\usepackage{amssymb}
\usepackage{bm}

\begin{document}

\title{Effect of the tensor force on the charge-exchange spin-dipole excitations of
$^{208}$Pb}

\author{C.L. Bai$^{1,2)}$, H.Q. Zhang$^{1,2)}$, H. Sagawa$^{3)}$, X.Z. Zhang$^{1)}$,
G. Col\`{o}$^{4)}$, and F.R. Xu$^{2)}$
}

\affiliation{$^{1)}$China Institute of Atomic Energy,
China}
\affiliation{
$^{2)}$School of Physics and State Key Laboratory of
Nuclear Physics and Technology, Peking University, China\\
$^{3)}$Center for Mathematics and Physics, University of Aizu, Aizu-Wakamatsu,
Fukushima 965-8560, Japan\\
$^{4)}$Dipartimento di Fisica, Universit$\grave{a}$ degli Studi di
Milano and INFN, Sezione di Milano, 20133 Milano, Italy
}


\begin{abstract}
The charge-exchange spin-dipole (SD) excitations of $^{208}$Pb are
studied by using a fully self-consistent Skyrme Hartree-Fock plus
Random Phase Approximation (HF+RPA) formalism which includes the
tensor interaction. It is found, for the first time, that the tensor
correlations have a unique, multipole-dependent effect on the SD
excitations, that is, they produce softening of 1$^-$ states, but
hardening of 0$^-$ and 2$^-$ states. This paves the way to a clear
assessment of the strength of the tensor terms. We compare our
results with  a recent measurement, showing that our choice of
tensor terms improves the agreement with experiment. The robustness
of our results is supported by the analytic form of the tensor
matrix elements.
\end{abstract}


\maketitle


The nuclear effective interactions like the zero-range Skyrme forces
have been quite successful to describe
many nuclear properties.
These forces are fitted
using empirical properties of uniform nuclear matter,
together with masses and charge radii of selected reference nuclei.
They describe
in a reasonable way the global trends of the ground-state
properties along the nuclear chart (binding energies, radii and
deformations).
Properties of excited states such as vibrations and rotations have been
studied successfully as well, allowing a large
amount of physical insight \cite{RMP,bulk}.
In the quest for a universal
{\em local} Energy Density
Functional (EDF) for nuclei, the Skyrme framework is
often used as a starting point.

While most of the Skyrme parameter sets which have been
widely used are purely central, many groups have recently
devoted attention to the role played by the zero-range tensor
terms that can be added
(see Refs.
\cite{Bro.06,Dob.06,Colo,Brink,Lesinski,Gra.07,Zou.08,Zal.08}).
This blooming of theoretical studies has followed the claim
by the authors of Ref.~\cite{Otsuka}, that the tensor force is
crucial for the understanding of the evolution of the
single-particle energies in exotic nuclei.

There exist different strategies to fix
the tensor part of the interaction. One can be
inspired by a bare or a G-matrix
interaction~\cite{Bro.06,Stancu}. Since the tensor force affects
the spin-orbit splittings as described below, another possibility is
to add it to  an existing Skyrme set and try to
reproduce at best the evolution of single-particle states
along isotopic or isotonic chains~\cite{Colo, Brink}. At present,
the most accurate and systematic way to produce effective interactions
with the tensor, has been a full variational procedure to
fit the tensor and the central terms on equal
footing~\cite{Lesinski}. All these attempts have produced results
which are not at all conclusive.


The studies based on a refit of all Skyrme parameters suffer
from the drawback that
the tensor force has a moderate effect on the ground
state quantities such as the total binding energies.
The single-particle energies do not lie within the EDF framework and
can be affected by correlations like particle-vibration coupling.
Then, we follow in this work the idea that
collective excitations (especially the spin-dependent
ones) may be a better candidate to
constrain the tensor force.

Recently, self-consistent HF+RPA schemes with tensor interactions have
been developed~\cite{Bai1,Bai2,Gang10}.
The Gamow-Teller (GT) and charge-exchange $1^{+}$ spin-quadrupole
(SQ) transitions in $^{90}$Zr and $^{208}$Pb have been studied in
Refs. \cite{Bai1,Bai2}, whereas the non charge-exchange multipole
responses of several magic nuclei have been calculated in
Ref. \cite{Gang10}.
In this letter, we study the charge-exchange spin-dipole (SD)
excitations of $^{208}$Pb, inspired by recent accurate measurements
\cite{Wakasa}.
Our specific goal is to find a clear, unambiguous signature
of the effective tensor force.
Spin-isospin collective modes have been instrumental for the
understanding of nuclear structure
since almost three decades~\cite{Osterfeld, Harakeh}.
In particular the total SD
strength distribution in $^{90}$Zr has been
measured~\cite{Wakasa1,Yako}
and this has allowed to extract conclusions on the
neutron radius~\cite{Sagawa1}. Neutron radii have been shown to
be strongly correlated with the features of the neutron
matter equation of state (EOS) which, in turn, has relevant
implications for neutron stars \cite{Furnstahl, Danielewicz, Yoshida}.
However, for the determination of the effect of the tensor
force which is strongly spin-dependent one needs to know
separately the
strength distributions of the $J^\pi=0^-$, $1^-$ and $2^-$ components.

The triplet-even and triplet-odd zero-range tensor terms of the Skyrme
force are expressed as~\cite{Stancu,Skyrme}
\begin{eqnarray}
V^{T}&=&\dfrac{T}{2}\{[(\mathbf{\sigma _{1}\cdot {k}^{\prime
})(\sigma _{2}\cdot {k}^{\prime })-\frac{1}{3}\left( \sigma
_{1}\cdot \sigma _{2}\right) {k}^{\prime 2}]\delta(r) }\nonumber\\
&+&\delta (\bf {r})[ (\bf{\sigma _{1}\cdot {k})(\sigma _{2}\cdot
{k})-\frac{1}{3}\left(
\sigma _{1}\cdot \sigma _{2}\right) {k}^{2}}] \}\nonumber\\
&+&\dfrac{U}{2}\{\left( \sigma _{1}\cdot \bf{k}^{\prime }\right)
\delta (\bf{r}) (\sigma _{2}\cdot \bf{k}) +\left( \sigma _{2}\cdot
\bf{k}^{\prime }\right) \delta (\bf{r})(\sigma _{1}\cdot
{k})\nonumber\\
&-&\frac{2}{3}\left[ (\bf{\sigma }_{1}\cdot \bf{\sigma }_{2})
\bf{k}^{\prime }\cdot \delta (\bf{r})\bf{k} \right]
\}.
\label{tensor}
\end{eqnarray}
In the above expression, the operator ${\bf k}=\left(\bf
\nabla_1-\bf \nabla_2\right)/2i$ acts on the right and ${\bf
k}^\prime=-\left(\bf \nabla_1^{\prime}-\bf
\nabla_2^{\prime}\right)/2i$ acts on the left. The coupling
constants $T$ and $U$ denote the strengths of the triplet-even and
triplet-odd tensor interactions, respectively.

The main effect of the tensor terms on the HF results is a
modification of the spin-orbit potential, which reads
\begin{eqnarray}
U_{S.O.}^{(q)}=\frac{W_0}{2r}(2\frac{d\rho_q}{dr}+
\frac{d\rho_{q\prime}}{dr})
+(\alpha\frac{J_q}{r}+\beta\frac{J_{q\prime}}{r}).
\label{spin_orbit}
\end{eqnarray}
In this expression, $q$=0(1) labels neutrons (protons).
$\rho_q$ are the densities while $J_q$ are the so-called
spin-orbit densities.
Their definitions can be found in
Refs.~\cite{Bro.06,Dob.06,Colo,Brink,Lesinski,Gra.07,Zou.08,Zal.08,Stancu}.
The first term in the r.h.s of Eq. (\ref{spin_orbit}) comes from the
Skyrme two-body spin-orbit interaction, whereas the second term
includes both a central exchange and a tensor contribution, that is,
$\alpha=\alpha_C+\alpha_T$ and $\beta=\beta_C+\beta_T$: their
complete expressions can be found e.g. in Ref. \cite{Colo,Zou.08}.
In this letter, we employ different Skyrme parameter sets which
include the tensor terms. The sets T$IJ$ have been introduced in
Ref.~\cite{Lesinski}.
The set SLy5+T$_w$ is a set in which the tensor terms are added perturbatively
on top of the existing force SLy5: the tensor parameters $T$ and $U$
are obtained by the low-$q$ limit the G-matrix calculations~\cite{Stancu}.
One should notice that the tensor part of SLy5+T$_w$ is different
from that of SLy5+T which was adopted in Ref.~\cite{Colo}.
We tried to use the set SLy5+T, but its effect on SD states is
larger and do not match the SD experimental data in
$^{208}$Pb~\cite{Wakasa}.
The values of $T$, $U$, $\alpha$ and $\beta$ for the
adopted interactions are listed in Table \ref{Para}.
\begin{table}[hbt] \centering
\caption{The tensor strength parameters $T$ and $U$ of Eq. \eqref{tensor} as
well as
the $\alpha$ and $\beta$ values of Eq. \eqref{spin_orbit}. All values are
in MeV$\cdot$fm$^5$.
\label{Para}}
\begin{tabular}{ccccccc}
\hline\hline Force & T & U & $\alpha_T$ & $\beta_T$ & $\alpha_c$ &
$\beta_c$ \\
\hline \hline
SLy5+T$_w$ & 820.0 & 323.4 & 134.76 & 238.2 & 80.2 & -48.9\\
T11 & 259.0 & -342.8 & -142.8 & -17.5 & 82.8 & -42.5\\
T22 & 356.0 & -217.5 & -90.6 & 28.9 & 90.6 & -28.9 \\
T32 & 613.1 & -2.3 & --96.5 & 79.5 & 96.5 & -19.5 \\
T43 & 590.6 & -147.5 & -61.5 & 92.3 & 121.5 & 27.7\\
T44 & 520.1 & 21.5 & 9.0 & 112.8 & 111.0 & 7.1 \\
\hline\hline
\end{tabular}
\end{table}

In our model,
we first solve the HF equations
in coordinate
space.
The unoccupied levels
are found by diagonalizing the mean field in
a harmonic oscillator basis (up to a maximum value of the
major quantum number $N_{max}=$
12).
We then perform
fully self-consistent RPA by including both the two-body spin-orbit
and tensor interactions.
It has been checked
that the adopted basis is large enough to make
the results stable.

The charge-exchange SD operator is defined as
\begin{eqnarray}
\hat{O}_{\pm
}^{\lambda }=\sum\limits_{i}t_{\pm
}^{i}r_{i}(Y_{1}^{i}\sigma^{i})_{\lambda}. \label{multipole}
\end{eqnarray}
The $n-$th energy weighted sum rules $m_n$ for the
$\lambda$-pole SD operator are defined as
\begin{eqnarray}
m_n^{\lambda}(t_{\pm})=\sum\limits_{i}E_i^n|\langle i|
\hat{O}_{\pm}^\lambda|0 \rangle|^2.
\end{eqnarray}
The model independent sum rule which is known to hold is
$m_0^{\lambda}(t_{-})
-m_0^{\lambda}(t_{+})$=
$\frac{2\lambda+1}{4\pi} (N \langle r^2 \rangle_n-Z \langle r^2
\rangle_p)$. This sum rule has been shown to
be fulfilled with
1\% accuracy by the numerical calculations.

\begin{table}[h] \centering
\caption{The SD sum rules $m_0$ and $m_1$ for $^{208}$Pb
with and without the tensor terms.
$\Delta E$ is the difference
between $m_1/m_0$ calculated with and without tensor.
\label{Table1}}
\begin{tabular}{c|c|ccc|ccc|r}
\hline\hline
   &  &  \multicolumn{3}{c|} {without tensor} & \multicolumn{3}{c|} {with tensor} & \\ \hline
  force & $\lambda^{\pi}$ &   $m_0$  &  $m_1$ & $m_1/m_0$ &  $m_0$  &  $m_1$ & $m_1/m_0$
& $\Delta E$ \\\hline \hline
     &  $0^-$   & 158.6 & 4718 & 29.7  & 171.9   & 5168   & 29.9 & 0.2\\
  SLy5  &  $1^-$ & 432.0 & 11746 & 27.2 & 440.0 & 10111 & 23.0 & $-$4.2\\
  +T$_w$   &      $2^-$   & 646.0 & 13742 & 21.3 & 657.4 & 14408 & 21.3 & 0\\ \cline{2-9}
     &  sum &   1236.4 & 30206 &  24.4 & 1269. & 29256 &23.0  & $-$1.4 \\
\hline
     &  $0^-$   &  158.6 & 4559 & 28.8 & 163.4 & 6074 & 37.2 & 8.4\\
  T11  &   $1^-$  & 434.5 & 11581 & 26.7 & 431.0 & 10376 & 24.1 & $-$2.6\\
     &   $2^-$   & 657.3 & 13710 & 21.2 & 646.2 & 14296 & 22.1  & 1.0\\\cline{2-9}
    &  sum  &   1240.  & 29850  &24.1 & 1241. & 30746 & 24.8 & 0.7 \\
  \hline
     &   $0^-$   &  157.6 & 4650 & 29.5 & 163.7 & 5943 & 36.3 & 6.8 \\
  T22  &   $1^-$   & 434.5 & 11771 & 27.1 & 433.9 & 10504 & 24.2 & $-$2.9 \\
     & $2^-$    & 645.9 & 13812 & 21.4 & 647.1   & 14327 & 22.1 & 0.7 \\\cline{2-9}
    &  sum  &  1238.0  &  30233  & 24.4 & 1244.7  & 30774 & 24.7 & 0.3\\
 \hline
     &   $0^-$   & 157.2 & 4698 & 29.9 & 166.9 & 6479 & 38.8 & 8.9 \\
  T32     &   $1^-$   & 435.3 & 11886 & 27.3 & 435.3 & 9928 & 22.7 & $-$4.6 \\
     &   $2^-$   & 646.1 & 13897 & 21.5 & 649.6 & 14619 & 22.5 & 1.0 \\\cline{2-9}
    &  sum &  1238.6  & 30481 &  24.6  &  1253.2  &  31026 & 24.8 & 0.2\\
\hline
     &  $0^-$   & 154.8 & 4693 & 30.3 & 164.0   & 6170   & 37.5 & 7.2 \\
   T43   &  $1^-$    &  440.3 & 12138 & 27.6 & 444.1 & 10366 & 23.3 & $-$4.3\\
      &     $2^-$    & 645.5 & 14067 & 21.8 & 649.4 & 14675 & 22.6 & 0.8 \\\cline{2-9}
    &  sum &  1241.  & 30898  & 24.9  &  1258.  & 31211  & 24.8 & $-$0.1\\
\hline
     &   $0^-$  & 155.6 & 4811 & 30.9  & 163.2 & 5637 & 34.5 & 3.6 \\
  T44     &  $1^-$  & 436.3 & 12174 & 27.9  & 440.6 & 10854 & 24.6 & $-$3.3\\
     &   $2^-$  & 664.2 & 14059 & 21.8 & 649 & 14406 & 22.2 & 0.4 \\\cline{2-9}
    &     sum & 1256.   &   31044 &  24.7  &  1253.  &  30897  &  24.7 & 0.0\\
  \hline\hline
\end{tabular}\thinspace
\end{table}

We performed two kinds of calculations.
In the first one, the tensor
terms are neither included in HF nor in RPA.
In the second one, the tensor terms are included both in HF and
in RPA.
 The terms containing the spin-orbit densities $J_q$ which arise from the
central momentum-dependent part of the Skyrme interaction are included in  all the following
HF and RPA calculations.
Previously, it had been found that the effect of tensor correlations in
HF is large for the Gamow-Teller mode and also, to some extent,
for some low-lying non charge-exchange excitations like
the first 2$^+$.
This is largely due to the fact that unperturbed GT transitions are
exactly those among spin-orbit partners. The unperturbed p-h energy
of the lowest  2$^+$ is  also determined by the  spin-orbit splitting.
On the other hand, this is not the case for the
SD excitations:
the average unperturbed energies are not much  affected by the spin-orbit
splittings since they are
  the 1$\hbar \omega$ type excitations.

\begin{figure}[h]
\includegraphics[scale=0.55,clip]{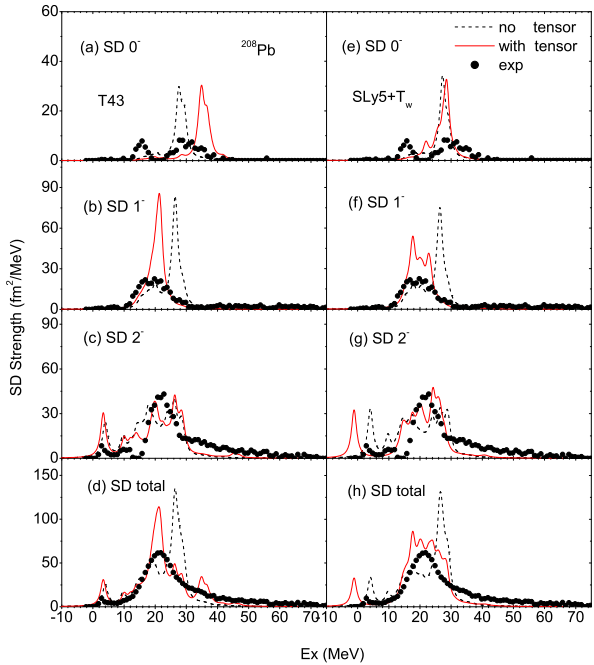}
\caption{(Color online) Charge-exchange SD$_-$ strength
distributions in $^{208}$Pb. In the panels (a), (b), and (c) the
RPA results obtained by employing the interaction
SLy5+T$_w$ for the multipoles 0$^-$ , 1$^-$,  2$^-$ are displayed.
In panel (d) we show the total strength
distribution. Panels (e), (f), (g) and (h) correspond
to similar results when the parameter set T43 is employed.
All these discrete RPA results have been smoothed by using a Lorentzian
averaging with a width of 2 MeV and compared with experimental
findings. The excitation energy is with respect to the ground state
of $^{208}$Bi. The experimental data are taken from Ref.~\cite{Wakasa}.
  See the text for details and discussion.
}
\label{fig1}
\end{figure}
The numerical
results of the HF+RPA calculations with the forces T43 and SLy5+T$_w$
are shown in Fig. \ref{fig1}.
They are compared with experimental
data obtained by Distorted Wave Impulse Approximation (DWIA) and
multipole decomposition analysis of the (p,n)
reaction \cite{Wakasa}. From Fig.\ref{fig1}(a) and (b) one can see that
in the case of the T43 interaction the main peaks of the 0$^-$ and
1$^-$ strength distributions are shifted upwards by about 7.5 MeV
and downwards by about 5 MeV, respectively,
due to the tensor correlations.
There are several 2$^-$ peaks (cf. Fig.\ref{fig1}(c)).
The peak at excitation energy $E_x\approx$17.7 MeV is moved upward by about 2 MeV by including
tensor, and comes close to an experimental peak, while another peak at
$E_x\approx$3.9 MeV is shifted downwards by about 0.6 MeV and is also
eventually close to the observed low energy peak.
For the total SD strength in Fig. \ref{fig1}(d), it is remarkable that
the main peak at 26 MeV is shifted to 21 MeV
when tensor is included, and this provides good agreement with the
experimental data.

In the same figure, the SD strength distributions in $^{208}$Pb
calculated by using the set SLy5+T$_w$ are also shown. From Fig.
\ref{fig1}(e), we see that the calculated 0$^-$ strength is
concentrated in one peak which is shifted upwards by about 1.3 MeV
by the tensor correlations. In Fig.\ref{fig1}(f), the RPA tensor
correlations move the 1$^-$ peak downwards and split it into three
peaks, in qualitative agreement with the bump-like experimental
strength. In the case of the 2$^-$ [Fig.\ref{fig1}(g)], the main
peaks in the high energy region are rather near to the experimental
main peak. Therefore, as shown in Fig.\ref{fig1}(h), the inclusion
of the tensor terms in HF+RPA can make the calculated main peak of
the total SD strength coincide with the main measured peak. However,
in low energy region the agreement is not good compared with the
experimental result in the case of SLy5+T$_w$ (due also to the fact
that with tensor a peak appears below the g.s. of the daughter
nucleus).
 The spin-orbit density gives contributions to the
p-h matrix elements of spin-dependent excitations~\cite{Bender}.
We find out in the case of
T22 parameter set that the two-body interaction generated
by the terms containing $J_q$ from the central exchange interaction
have a much smaller effect on the SD excitations than
the two-body tensor interaction~\cite{Bai11}.


We would like at this stage to obtain a better understanding of this
peculiar role of tensor interactions.
The diagonal matrix element of the triplet-even (TE) term on a state
with multipolarity $\lambda$ can be expressed as \cite{Gang10}
\begin{eqnarray}
V^{(\lambda)}_{TE} & = &
\frac{5T}{4}
\sum_{\ell,k,k'}
\frac{(-)^{k+k'+\lambda+\ell+1}\hat{k} \hat{k'}}{2\lambda +1}
\left\{ \begin{array}{ccc} k & k' & 2 \\
1 & 1 & \ell \end{array} \right\} \nonumber  \\
& \times &
\left\{ \begin{array}{ccc} 1 & 1 & 2  \\
k' & k & \lambda \end{array} \right\}
\langle p  \vert\vert \hat O_{k',\lambda} \vert\vert h \rangle
\langle p  \vert\vert \hat O_{k,\lambda}  \vert\vert h  \rangle ^*,
\label{eq:V-te}
\end{eqnarray}
in terms of the reduced matrix elements of the operator
$\hat O_{k,\lambda}= \sum_i [ \sigma_i \otimes (
{\bf \nabla}_i \otimes Y_\ell(i) )^{(k)}
]^{(\lambda)}$ and $6j$ symbols. In Eq. \eqref{eq:V-te}, the notation
$\hat k\equiv\sqrt{2k+1}$ is used.
For the SD excitations, taking $\ell=0$ and $k$=$k'$=1,
Eq. \eqref{eq:V-te} gives
\begin{equation}
V^{(\lambda)}_{TE} = -
\frac{5}{12}T \left\{ \begin{array}{c}
 1 \\
-1/6 \\
  1/50 \end{array} \right\}
|\langle p  \vert\vert \hat O_{1,\lambda} \vert\vert h \rangle|^2 \,\,\, for \,\,\,  \lambda=\left\{ \begin{array}{c}
                                                    0^- \\
                                                    1^- \\
                                                    2^- \end{array} \right\}.
\label{eq:V-TE}
\end{equation}
The TO tensor part is also expressed in a similar way as
\begin{equation}
V^{(\lambda)}_{TO} = -\frac{5}{12}U \left\{ \begin{array}{c}
1 \\
-1/6 \\
1/50   \end{array} \right\}
|\langle p  \vert \vert \hat O_{1,\lambda} \vert \vert  h \rangle|^2 \,\,\, for \,\,\,  \lambda=\left\{ \begin{array}{c}
                                                    0^- \\
                                                    1^- \\
                                                    2^- \end{array} \right\}.
\label{eq:V-TO}
\end{equation}
We can see in Eqs. \eqref{eq:V-TE} and  \eqref{eq:V-TO}
 that the diagonal
  p-h matrix element in the 0$^-$ case is the largest, and
that for 1$^-$ is the next. The effect on $2^-$ is rather small.
It should be noticed  that
these relative strengths of the Skyrme tensor interactions on each multipole are similar to
those obtained from the finite-range tensor interactions both
 in magnitude and in sign \cite{SB84}. We can sum the TE and TO direct matrix elements as
\begin{equation}
V^{(\lambda)}_{T}=V^{(\lambda)}_{TE}+V^{(\lambda)}_{TO}\equiv a_{\lambda}T+b_{\lambda}U.
\end{equation}
The proper antisymmetrization is easy to obtain for contact interactions
and gives, in the isovector channel,
\begin{equation}
V^{(\lambda)}_{T,AS}=[-\frac{1}{2}a_{\lambda}T+\frac{1}{2}b_{\lambda}U ]\langle \tau_1\cdot\tau_2 \rangle.
\label{eq:tensor-a}
\end{equation}
Since the coupling constant $T$
is
positive for the interactions we considered, $V^{(\lambda)}_{TE}$
is repulsive for the $0^-$ and $2^-$ case, while it is attractive for $1^-$ .
The $V^{(\lambda)}_{TO}$ part may contribute with
  the same sign as the $V^{(\lambda)}_{TE}$ one if the value of $U$ is negative.
For the T$IJ$ family, the value of $U$ is negative or small
positive,
so that the
$V^{(\lambda)}_{TO}$ contributions have the same multipole dependence
  or almost negligible.
All together, the tensor correlations are strongly repulsive
  for 0$^-$ and
weakly repulsive for 2$^-$  in general.
For
$1^-$, the net effect will be  attractive.
For SLy5+T$_w$, the value of $U$ is positive and will give opposite
contributions to those of $T$.  However, the $T$ value is much larger
than the value of $U$ so that the same argument given for the T$IJ$ family
will hold.
One can see from Table \ref{Table1} that Eq. \eqref{eq:tensor-a} provides
a very effective guideline for interpreting the numerical results of
microscopic RPA.
In the same Table we also provide values of the sum rules $m_0$ and
$m_1$ for the different multipoles and effective forces.

In the parameter sets T43 and T44 used for Table I, the spin-orbit strengths
 are  larger than in the other parameter sets.  However, these large spin-orbit
strengths cancel  substantially with  the effect of the terms
$\alpha$ and $\beta$ of Eq. (2), and the net results are quite similar for the
single particle spectra of the parameter sets in Table I.


In summary, we have studied the effect of tensor interactions on the
charge-exchange
$t_-$ SD excitations of $^{208}$Pb by using the self-consistent HF+RPA model.
We have demonstrated clearly for the first time that tensor correlations
have a specific multipole dependence,
that is, they produce a strong hardening effect on the 0$^-$mode and a
softening effect on the 1$^-$ mode. A weak hardening effect is also
observed on the 2$^-$ mode.
The characteristic effect of the tensor force was further examined
by using the analytic formulas based on the multipole expansion of
the contact tensor interaction. It is found that both the TE and TO
tensor
interactions do  provide the characteristic multipole dependent
effect on the SD excitations.
These effects are shown to be
robust irrespective
of the adopted Skyrme force with tensor interactions.
Our calculated results are compared with recent SD excitation spectra
obtained in the $(p,n)$ reaction on the $^{208}$Pb target.
The softening and the hardening effects produced by
tensor correlations on the 1$^-$ and 2$^-$ modes are confirmed
by comparing the experimental data and the calculated results with and
without tensor correlations.
Consequently, from the study of SD excitations,
 we are able to give a clear constraint on the effective
tensor force which was missing so far.

We would like to thank T. Wakasa for useful discussions and
providing us his data prior to
the publication. We would like to thank also
Hide Sakai and
Witek Nazarewicz for enlightening discussions.
This work is supported by the National Natural Science Foundation
of China under Grant Nos. 10875172, 10275092 and 10675169, the
National Key Basic Research Program of China under Grant
No. 2007CB815000. This research was supported in part by the
Project of Knowledge Innovation Program (PKIP) of Chinese Academy
of Sciences, Grant No. KJCX2.YW.W10,
and the Japanese Ministry of Education, Culture,Sports, Science
and Technology by Grant-in-Aid
for Scientific Research under the program number (C(2))20540277.


\end{document}